\documentclass[aps,prl,twocolumn,showpacs,groupedaddress]{revtex4}
\usepackage{amsmath}
\usepackage{amssymb}
\usepackage{epsfig}
\usepackage{graphicx,amsmath}
\usepackage{calc}
\usepackage{bm}
\begin{document}
\title{The effective electron mass in high-mobility SiGe/Si/SiGe quantum wells}
\author{M. Yu. Melnikov, A. A. Shashkin, and V. T. Dolgopolov}
\affiliation{Institute of Solid State Physics, Chernogolovka, Moscow District 142432, Russia}
\author{S. V. Kravchenko}
\affiliation{Physics Department, Northeastern University, Boston, Massachusetts 02115, USA}
\author{S.-H. Huang and C. W. Liu}
\affiliation{Department of Electrical Engineering and Graduate Institute of Electronics Engineering, National Taiwan University, Taipei 106, Taiwan, and\\ National Nano Device Laboratories, Hsinchu 300, Taiwan}
\date{\today}

\begin{abstract}
The effective mass, $m^*$, of the electrons confined in high-mobility SiGe/Si/SiGe quantum wells has been measured by the analysis of the temperature dependence of the Shubnikov-de~Haas oscillations. In the accessible range of electron densities, $n_s$, the effective mass has been found to grow with decreasing $n_s$, obeying the relation $m^*/m_b=n_s/(n_s-n_c)$, where $m_b$ is the electron band mass and $n_c\approx 0.54\times 10^{11}$~cm$^{-2}$. In samples with maximum mobilities ranging between 90 and 220 m$^2$/Vs, the dependence of the effective mass on the electron density has been found to be identical suggesting that the effective mass is disorder-independent, at least in the most perfect samples.
\end{abstract}

\pacs{71.10.Hf, 71.27.+a, 71.10.Ay}
\maketitle

It has been reliably established that the effective electron mass in two-dimensional (2D) electron systems in silicon metal-oxide-semiconductor field-effect transistors (MOSFETs) grows dramatically with decreasing electron density and, as a consequence, with the increasing strength of interactions. Subject to the validity of the quasiparticle description in the strongly-correlated electron systems, the most direct way of measuring the effective mass has been the analysis of the temperature dependence of the amplitude of the Shubnikov-de~Haas oscillations \cite{shashkin2003,pudalov2002,shashkin2007}. This method yields the effective mass of the quasiparticles near the Fermi level defined as $m^*=p_F/v_F$, where $p_F$ and $v_F$ are Fermi momentum and Fermi velocity, correspondingly. A certain weakness of this method is that the use of the Lifshitz-Kosevich formula \cite{lifshitz1955} was not fully justified. The formula assumes the existence of a large number of Landau levels below the Fermi level, while in the experiments mentioned above, in the range of parameters where the strong increase of $m^*$ has been observed, this number did not exceed three.

The dramatic increase of the effective mass has also been confirmed in a variety of other experiments. One of the methods was based on the fact that the temperature dependence of the elastic mean-free time, $\tau$, is determined by the same parameter, $p_F/v_F$, as the amplitude of the Shubnikov-de~Haas oscillations. The analysis of the low-$n_s$  linear temperature dependence of $\tau$ in the spirit of theories \cite{gold1986,zala2001} has yielded \cite{shashkin2002,pudalov2003} the behavior of the effective mass similar to that obtained from the studies of the Shubnikov-de~Haas oscillations. Yet another confirmation of the interaction-induced increase of the effective mass has been obtained by studies of the full spin polarization of the electrons in a magnetic field, parallel to the 2D plane \cite{shashkin2005,shashkin2001}, and by magnetization measurements in the ballistic regime \cite{prus2003,shashkin2006}. It should be noted, however, that the latter two methods give a differently defined effective mass. In interpreting the experimental results, the energy of the fully or partially spin-polarized electron system was compared to the energy of the spin-unpolarized electron system and it was assumed that either Fermi energy can be written as $p_F^2/2m^*$, where the Fermi momenta for spin-polarized and spin-unpolarized electrons are different, but the effective masses are equal. In the range of electron densities down to $\approx 1\times 10^{11}$~cm$^{-2}$, the results obtained are in good agreement with the results obtained by other methods, which indicates that the effective mass is greatly enhanced, while the Lande $g$ factor stays close to its value in bulk silicon (see Ref.~\cite{shashkin2005}).

Finally, we should mention measurements of the thermopower \cite{mokashi2012}, which, although also not completely assumption-free, yielded a more than an order of magnitude growth of the effective mass $m^*=p_F/v_F$ with decreasing electron density, in good agreement with the other data.

However, there exists a certain disagreement in the interpretation of the experimental data. In the majority of papers \cite{shashkin2003,shashkin2007,shashkin2002,shashkin2005,shashkin2001,shashkin2006,mokashi2012, vitkalov2002}, the conclusion was made that the effective electron mass in silicon MOSFETs behaves critically tending to infinity at a finite electron density, $n_c$. This was based on the extrapolation of the experimental data obtained in the ``good'' metallic regime $\sigma>e^2/h$, where the effective mass obeyed the equation
\begin{equation}
\frac{m^*}{m_b}=\frac{n_s}{n_s-n_c}.
\label{mass}
\end{equation}

In other publications, some doubts in this interpretation were expressed (for review, see Ref.~\cite{giuliani2004}). There also exist experiments in which the influence of the disorder potential (at least, a strong one) on the experimental results was reported \cite{vitkalov2002,pudalov20021}.

There is no consensus in the theoretical predictions either. According to Refs.~\cite{amusia2014,dolgopolov2012,amaricci2010}, critical behavior of the effective mass is an intrinsic property of any strongly-correlated 2D system, while in Refs.~\cite{fleury,senatore}, a conclusion has been made that in a clean two-valley electron system, similar to the one in low-disorder silicon MOSFETs, critical behavior is impossible and may exist only as a consequence of the existence of a disorder potential \cite{punnoose2005}. To answer these questions, experiments must be conducted on 2D electron systems of much higher quality than the previously studied Si MOSFETs, in which the maximum electron mobilities did not exceed 3 m$^2$/Vs.

The aim of the present work is to measure the effective electron mass in extremely low-disorder SiGe/Si/SiGe quantum wells with mobilities up to 220 m$^2$/Vs, {\it i.e.}, almost two orders of magnitude higher than those in the best of previously studied silicon MOSFETs.

\begin{figure}
\includegraphics[width=8.5cm]{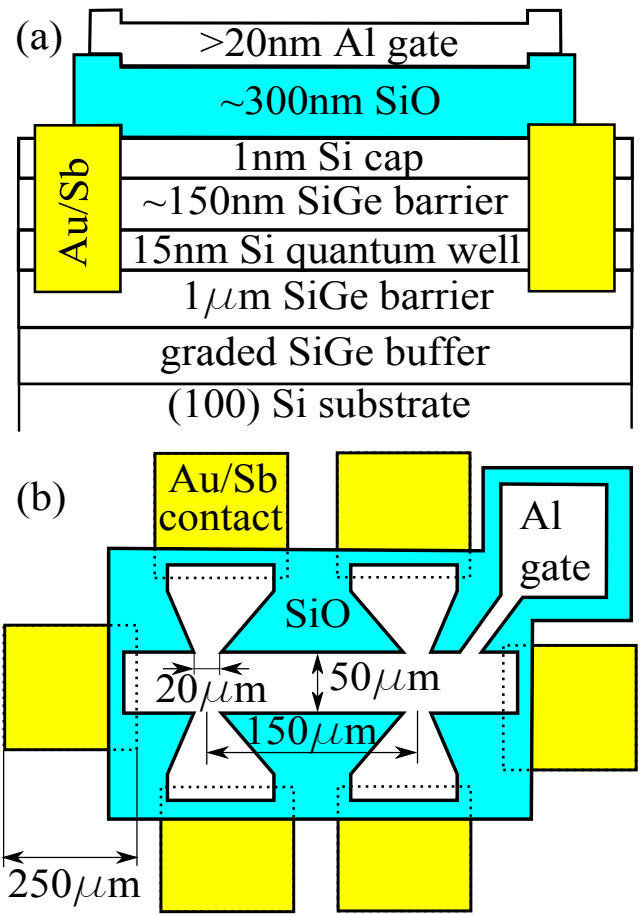}
\caption{\label{fig:scheme} (color online). (a) Band diagram of the sample. (b) Schematic top view of the sample.}
\end{figure}

To make a sample, we used a SiGe/Si/SiGe quantum well grown in an ultrahigh-vacuum chemical-vapor-deposition (UHVCVD)  (for details, see Refs.~\cite{lu2009,erratum}). Approximately 15 nm wide silicon quantum well is sandwiched between SiGe potential barriers (Fig.~\ref{fig:scheme}~(a)). The samples were patterned in Hall-bar shapes using standard photo-lithography (Fig.~\ref{fig:scheme}~(b)) on two different pieces, SiGe1 and SiGe2, of the same wafer. As the first step, electric contacts to the 2D layer were made. They consisted of AuSb alloy, deposited in a thermal evaporator in vacuum and annealed. Then, approximately 300 nm thick layer of SiO was deposited in a thermal evaporator and a $>20$ nm thick Al gate was deposited on top of SiO. The fabrication procedures of SiGe1 and SiGe2 differed significantly only in the way of how the electric contacts to the 2D layer were made. In case of SiGe2, after depositing approximately 350 nm Au$_{0.99}$Sb$_{0.01}$, the contacts were annealed for 5 minutes in N$_2$ atmosphere at 440$^\circ$C (the procedure used in Refs.~\cite{lu2009,erratum}). In case of SiGe1, first about 30 nm of Sb and then 230 nm of Au were deposited, and then annealing was made by an electric spark which was produced by touching the contact with a metallic needle while the second metallic needle, connected with the first one by a charged capacitor, was pressed to the Au/Sb surface. Several of such annealing were done for each of the contact pads along the edge, where the Al gate was subsequently deposited.

When a positive voltage $V_g>V_{th}\approx0$ is applied to the gate, an $\approx15$ nm wide electron system is formed in a Si (100) quantum well approximately 150 nm below the SiO layer. It is expected that the properties of such a system (the band electron mass, $g$-factor, two-valley spectrum) are identical with those of the 2D system in Si MOSFETs, with the exception of the characteristic energy of the electron-electron interactions. The latter is expected to be somewhat weaker than that in Si MOSFETs because of a greater average dielectric constant in SiGe/Si/SiGe ($\sim12$ compared to 7.7 in Si MOSFETs).

We have studied four samples of the same geometry, two of SiGe1 type and two of SiGe2 type, in an Oxford TLM-400 dilution refrigerator in a temperature range 0.05 -- 1.2 K. To measure the resistance, a standard four-terminal lock-in technique was used in a frequency range 1 -- 11 Hz; the applied currents varied in the range 0.5 -- 4 nA. We faced two main problems with the measurements. First, the ohmic contacts to the 2D layer became highly resistive already at relatively high electron densities $n_s\lesssim1.2$ -- $1.6\times10^{11}$ cm$^{-2}$ (depending on the sample) and disappeared altogether at yet lower densities. The results presented in this paper were obtained when the contact resistance was in the range between 0.5 and 100 k$\Omega$; a preamplifier with the input resistance of 100 M$\Omega$ was used to minimize the effect of the contact resistance. Second, it took rather long time for the electron density to stabilize after the gate voltage was changed: about two hours for SiGe1 type of samples and less than half an hour for SiGe2 type. The electron density was carefully monitored during the measurements. In both types of samples, if a high density above $2.4\times10^{11}$~cm$^{-2}$ was initially set by the gate voltage, after a few hours it would ultimately reduce to $n_s\approx2.4\times10^{11}$ cm$^{-2}$. The same effect was reported in Ref.~\cite{lu2011} where a similar maximum electron density of $2.7\times10^{11}$ cm$^{-2}$ was reported. The saturation of the electron density was explained by a tunneling of the electrons through the SiGe barrier at high gate voltages.

At $T=0.05$~K and $n_s=1.2$ -- $2.4\times10^{11}$ cm$^{-2}$, the maximum electron mobility varied between 90 and 220 m$^2$/Vs for the four samples studied. The threshold voltage $V_{th}\approx 0$ was determined by extrapolating the linear dependence of the stabilized electron density on the gate voltage. In contrast, a rather high threshold voltage $V_g=5.25$ V was reported in Refs.~\cite{lu2009,erratum} where Al$_2$O$_3$ was used as a dielectric between the structure and the gate. According to the authors, this might be due to the influence of the interface between the Al$_2$O$_3$ layer and the heterostructure.

\begin{figure}
\includegraphics[width=8.5cm]{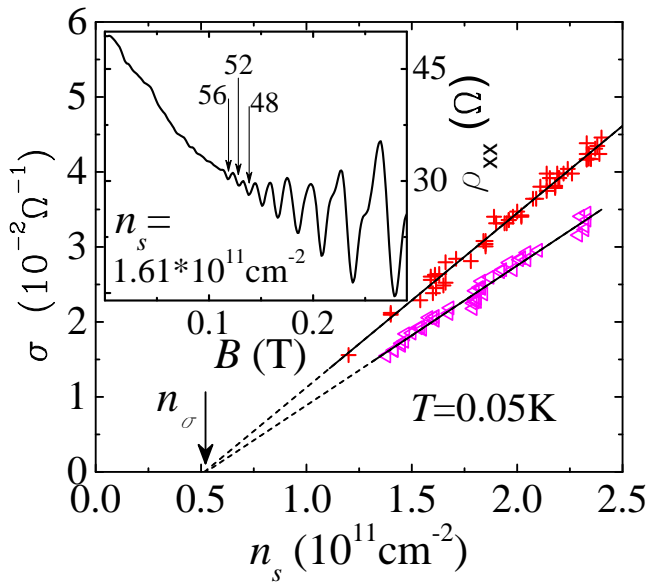}
\caption{\label{fig2} (color online). Conductivity as a function of $n_s$ for SiGe1-I (crosses) and SiGe1-II (triangles). The solid lines are obtained by the least mean square method. Being extrapolated to $\sigma=0$, they intersect with the $x$-axis at $n_\sigma\approx 0.52\times10^{11}$ cm$^{-2}$. The inset shows the longitudinal resistivity as a function of the perpendicular magnetic field for the sample SiGe1-II at $T=0.05$~K. Arrows show the positions of the minima of the oscillations, calculated for the filling factors indicated for $n_s=1.61\times10^{11}$ cm$^{-2}$.}
\end{figure}

The conductivity $\sigma(n_s)$ in zero magnetic field $B=0$ for two SiGe1 samples is shown in Fig.~\ref{fig2}. For each of the samples, the data were obtained in several cool-downs from room to helium temperature. The electron density was determined from the low-field Shubnikov-de~Haas oscillations (see the inset to Fig.~\ref{fig2}). Electron densities determined from the Hall effect were found to be the same within 10\%.

Both sets of data are described well by linear dependences $\sigma\propto n_s-n_\sigma$ where $n_\sigma\approx 0.52\times10^{11}$ cm$^{-2}$. We do not show the experimental data for the higher-mobility SiGe2 because they are more sparse and were obtained in a narrower region of densities. Within our accuracy, the data for SiGe2 extrapolate to the same $n_\sigma$ as the data for SiGe1.

The longitudinal resistivity $\rho_{xx}$ as a function of the perpendicular magnetic field is shown in the inset to Fig.~\ref{fig2}. The filling factors $\nu=n_s hc/eB$, corresponding to the minima of the Shubnikov-de~Haas oscillations, are factors of 4 in agreement with the existence of the two-fold spin and valley degeneracies of the Landau levels. The quantum oscillations start at a magnetic field of 0.1~T. This allows one to estimate the ``quantum'' mobility to be of order 10 m$^2$/Vs. However, the value of the mobility, calculated from the conductivity data, is an order of magnitude higher. It means that the transport relaxation time in our samples is an order of magnitude longer than the quantum time $\tau_q$ responsible for the width of the Landau levels. Similar an-order-of-magnitude difference between transport and quantum relaxation times has been observed on all four samples at all densities. This points to the predominantly small-angle scattering well known in the high-mobility heterostructures.

\begin{figure}
\includegraphics[width=8.5cm]{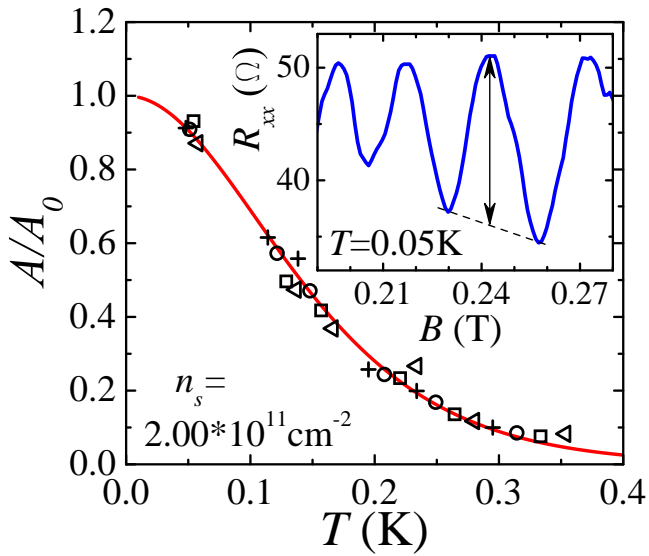}
\caption{\label{fig3} (color online). Temperature dependence of the amplitude of the Shubnikov-de~Haas oscillations in weak magnetic fields $B_1=0.260$~T (crosses), $B_2=0.244$~T (circles), $B_3=0.230$~T (squares), and $B_4=0.217$~T (triangles). Values of $T$ for $B_2$, $B_3$, and $B_4$ are multiplied by $B_1/B_2$, $B_1/B_3$, and $B_1/B_4$, correspondingly.  The inset shows the corresponding oscillations of the magnetoresistance. The curve is obtained by averaging five magnetoresistance traces. The arrow shows the double amplitude of the oscillation corresponding to $B_2$.}
\end{figure}

An example of the temperature dependence of the amplitude $A$ of the Shubnikov-de~Haas oscillations in a weak magnetic field is shown in Fig.~\ref{fig3}. All the temperature dependences were obtained in the regime where the Landau levels are four-fold. The amplitudes were determined as ratios of the half of the height of the oscillation (as shown in the inset to Fig.~\ref{fig3}) to the average resistance $R_0$. To determine the effective mass, we used the Lifshitz-Kosevich formula
\begin{equation}
A(T)=A_0\frac{2\pi^2 k_B T/\hbar\omega_c}{\text{sinh}(2\pi^2 k_B T/\hbar\omega_c)},
\label{LK}
\end{equation}
where
\begin{equation}
A_0=4\; \text{exp}(-2\pi^2  k_B T_D/\hbar\omega_c),
\end{equation}
$\omega_c=eB/m^*c$ is the cyclotron frequency, and $T_D=\hbar/2\pi k_B\tau_q$ is the Dingle temperature. In principle, a possible temperature dependence of $\tau_q$ may influence the suppression of the oscillations with temperature. However, in our experiments, possible corrections to the effective mass due to the temperature dependence of $\tau_q$ are within experimental uncertainty and do not exceed 10\%. Note that the measured amplitude of the oscillations follows the calculated curve down to the lowest temperatures thus confirming that the electrons were in good thermal contact with the crystal and the helium bath and were not overheated. Note also that Eq.~(\ref{LK}) was obtained for the case $A\ll1$. In spite of this, it describes our data well even when the amplitude reaches 25\% at the lowest temperature.

\begin{figure}
\includegraphics[width=8.5cm]{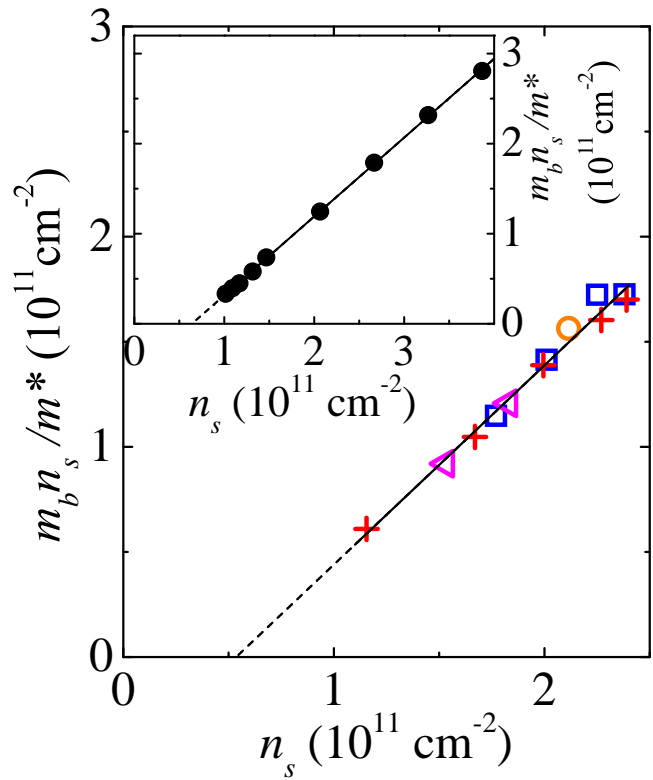}
\caption{\label{fig4} (color online). Dependence of $m_b n_s/m^*$ on $n_s$ for four samples: SiGe1-I (crosses),  SiGe1-II (triangles),  SiGe2-I (squares), and  SiGe2-II (a circle.) Linear approximation is shown by the solid line, and the extrapolation to the $x$-axis is depicted by the dashed line. The inset shows $m_b n_s/m^*$ {\it vs.} $n_s$ in Si MOSFETs \cite{shashkin2003}.}
\end{figure}

To compare the behavior of the effective mass with Eq.~(\ref{mass}), in Fig.~\ref{fig4} we show $m_b n_s/m^*$, the inverse effective mass multiplied by the electron density and by the electron band mass, {\it vs.} $n_s$ (the band mass in silicon is $0.19\; m_e$ where $m_e$ is the free electron mass). For all four samples, all the experimental points form a single line and are described well by the linear dependence $m_b n_s/m^*\propto n_s-n_c$. The line, obtained by the least mean square method, extrapolates to zero at $n_c\approx 0.54\times10^{11}$ cm$^{-2}$ within the experimental uncertainty of about 10\%. As seen from the figure, the slope of the line is very close to unity, {\it i.e.}, the coefficient in Eq.~(\ref{mass}) is close to the band mass of the electrons $0.19\; m_e$ in our samples, in agreement with Refs.~\cite{dolgopolov2012,amaricci2010}.

For comparison, in the inset of Fig.~\ref{fig4} we show the data obtained by the same method on silicon MOSFETs \cite{shashkin2003}. The linear fit intercepts the $x$-axis at a density $\approx 0.64\times 10^{11}$~cm$^{-2}$ which is somewhat lower than the critical density determined using more accurate methods \cite{mokashi2012}.

It is well known that the effective electron mass, defined as $p_F/v_F$, enters the equation describing the linear temperature-dependent corrections to the conductivity in the ballistic regime \cite{gold1986,zala2001}:
\begin{equation}
\sigma(T)=\sigma_0\left(1-\alpha\frac{k_BT}{E_F}\right),\label{linear}
\end{equation}
where $E_F=p_F^2/2m^*$ is the Fermi energy and $\sigma_0$ is the conductivity extrapolated to $T=0$. Below we will assume, in agreement with the results of Ref.~\cite{shashkin2002}, that the coefficient $\alpha$ does not depend on $n_s$.

\begin{figure}
\includegraphics[width=8.5cm]{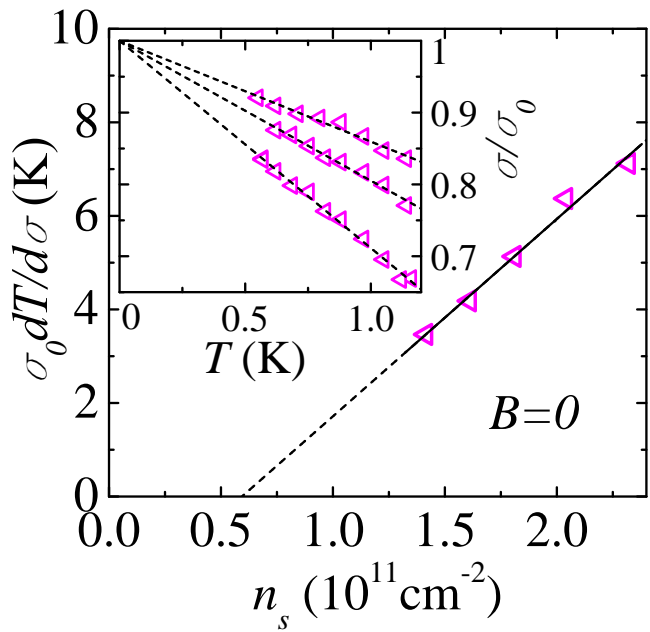}
\caption{\label{fig5} (color online). $\sigma_0dT/d\sigma$ as a function of $n_s$ for SiGe1-II (triangles). The solid line is a linear approximation and the dashed line is its extrapolation to the $x$-axis. The inset shows normalized conductivities {\it vs.} temperature for $n_s=1.42$, 1.81, and $2.32\times10^{11}$ cm$^{-2}$ from bottom to top. The dashed lines in the inset are linear approximations.}
\end{figure}

In the inset to Fig.~\ref{fig5}, we show the temperature dependences of the normalized conductivity of the sample SiGe1-II for three electron densities in the temperature range where the $\sigma(T)$ dependences are linear, {\it i.e.}, between 0.5 and 1.2 K. As follows from Eq.~(\ref{linear}), the product $\sigma_0dT/d\sigma$ is proportional to $n_s/m^*$. In the main panel of Fig.~\ref{fig5}, $\sigma_0dT/d\sigma$ is plotted as a function of $n_s$. The dependence is indeed linear once again confirming Eq.~(\ref{mass}). The extrapolation to zero yields, within our accuracy, the same critical density $n_c$ for all samples tested.

Comparison of the experimental data plotted in Figs.~\ref{fig2} and \ref{fig4} in the range of electron densities used ($n_s=1.2$ -- $2.4\times10^{11}$ cm$^{-2}$) suggests that the dependences are proportional to each other due to $n_\sigma\approx n_c$ and, therefore, the transport relaxation time is independent of $n_s$. Earlier the same conclusion was made based on the results obtained on high-mobility Si MOSFETs \cite{shashkin2006b}.

The experimental data for the effective mass are identical for all four samples measured in this work in spite of more than two-fold difference in mobilities. Obviously, this confirms that the experiments were done in the ``clean limit'' and the results are not influenced by the disorder potential.

In the range of electron densities used in this paper, the effective mass follows Eq.~(\ref{mass}), which suggests the divergence of $m^*$ at a finite electron density $n_c$. As shown above, the same conclusion indirectly follows from Fig.~\ref{fig5}. However, based on the data, one cannot make firm conclusions about the true critical behavior of the effective mass because the extrapolation $1/m^*\rightarrow0$ is made from electron densities too far from the critical point. Furthermore, a naive estimate of the critical electron density expected in this system yields somewhat lower values than $n_c$ obtained in our experiments. Indeed, it seems natural to assume that in all clean electron systems the divergence of the effective mass happens at the same value of the interaction parameter, {\it i.e.}, the ratio between the Coulomb and Fermi energies $r_s=E_c/E_F\propto 1/\epsilon n_s^{1/2}$ (here $\epsilon$ is the average dielectric constant). Therefore, to reach the same values of $r_s$, electron densities must be lower by the factor of $(\epsilon_{\text{SiGe}}/\epsilon_{\text{Si MOSFET}})^2$ in SiGe/Si/SiGe quantum wells compared to those in Si MOSFETs. Since in Si MOSFETs the critical density $n_c\approx 0.8\times10^{11}$ cm$^{-2}$, one would expect it to be about $0.33\times10^{11}$ cm$^{-2}$ in SiGe/Si/SiGe quantum wells, well below $n_c\approx 0.54\times10^{11}$ cm$^{-2}$ found in our experiments. Obviously, to make definite conclusions about the divergence of the effective mass in SiGe/Si/SiGe quantum wells, the samples must be further improved; in particular, contacts should be modified to allow studies of the electron densities well below $10^{11}$ cm$^{-2}$.

In summary, the effective mass of strongly-correlated electrons in ultra-low-disorder SiGe/Si/SiGe quantum wells has been measured by the analysis of the temperature dependence of the Shubnikov-de~Haas oscillations. The effective mass has been found to be disorder-independent and to grow with the strength of the electron-electron interactions in a way similar to that in low-disorder Si MOSFETs. However, to reliably establish the critical behavior of the effective mass and its possible divergence at a critical electron density, measurements should be conducted at much lower values of  $n_s$ --- something that is currently impossible due to the high contact resistance at $n_s\lesssim 1.2\times 10^{11}$ cm$^{-2}$. Work on improving contacts is in progress.

We gratefully acknowledge discussions with G.M. Minkov. This work was supported by RFBR 12-02-00272 and 13-02-00095, RAS, and the Russian Ministry of Sciences. SVK was supported by NSF Grant 1309337 and BSF Grant 2012210.

\end{document}